\documentclass[12pt]{article}
\usepackage{epsfig}
\usepackage{txfonts}
\usepackage{natbib}

\def\I{\,{\sc i}}
\def\II{\,{\sc ii}}
\def\III{\,{\sc iii}}

\begin{document}

\centerline{\Large\bf The 2006 hot phase of Romano's star (GR 290)
in M~33\footnote{Based on observations collected with the 1.52 m
Cassini telescope of the Bologna Astronomical Observatory} }

\bigskip

\centerline{ R. F. Viotti (1),
  S. Galleti (2),  R. Gualandi (2), F. Montagni  (3)}
\medskip
\centerline{
 V. F. Polcaro (1), C. Rossi (3) and L. Norci (4)
 }

%R. F. Viotti, S. Galleti,  R. Gualandi, F. Montagni, V. F. Polcaro, C. Rossi and L. Norci 

%
%\offprints{\\R. F. Viotti, email: roberto.viotti@iasf-roma.inaf.it}

\bigskip

(1) INAF, Istituto di Astrofisica Spaziale e Fisica Cosmica di Roma,
via Fosso del Cavaliere 100, 00133 Roma, Italy

\bigskip (2) INAF, Bologna Astronomical Observatory, Loiano Observing
Station, Loiano, Italy

\bigskip (3) Dipartimento di Fisica, Universit\`a di Roma La
Sapienza, Pz.le Aldo Moro 3, 00185 Roma, Italy

\bigskip(4) School of Physical Sciences and NCPST, Dublin City
University, Glasnevin, Dublin 9, Ireland

\bigskip

\centerline{ A\&A Letters, Received 22 December 2006  / Accepted 28
January 2007}

\bigskip
\bigskip

%======================   ABSTRACT   ==========================

\centerline{\large\bf ABSTRACT}

\bigskip

Understanding the nature of the instabilities of LBVs is important to understand
the late evolutionary stages of very massive stars.

 We investigate the long term, S Dor-type variability of the
luminous blue variable GR290 (Romano's star) in M33, and
its 2006 minimum phase.

New spectroscopic and photometric data taken in November
and December
2006 were employed in conjunction with
already published data on GR290 to derive
the physical structure of GR290 in different phases
and the time scale of the variability.

We find that by the end of 2006,
GR~290 had reached the deepest visual minimum
so far recorded.
Its present spectrum resembles closely
that of the Of/WN9 stars, and is the hottest so far
recorded in this star (and in any LBV as well),
while its visual brightness decreased by about 1.4 mag.

This first spectroscopic record of GR290 during a minimum phase
confirms that, similarly to AG Car and other LBVs, the star is subject to ample
S Dor-type variations, being hotter at minimum, suggesting that
the variations take place at constant bolometric luminosity.

{\bf Keywords}: stars: evolution --
        stars: variable --
        stars: individual: GR~290 --
        stars: hypergiant --
        galaxies: individual: M~33

%\titlerunning{Hot phase of GR 290}
%\authorrunning{R. F. Viotti et al.}
%\maketitle

%=======================   SECTION  1   =====================

\section{Introduction and observations} \label{sec:Intr}

Romano's star (GR~290) is one of the most luminous hot variable
stars in M 33 (Viotti et al. 2006). Its historical light curve is
characterised by ample, long--term variations between 16.7 and 18.1
in the $B$--band (Romano 1978, Kurtev et al. 2001). This behaviour
is reminiscent of the long--term photometric variations of the S
Doradus variable stars that are generally accompanied by
colour--index variations, the stars being bluer when fainter in the
visual. The spectroscopic monitoring of a number of LBVs confirmed
that during minimum they display a hotter spectral type. For
instance AG Car changes from a Be/Ae spectral type at maximum visual
magnitude, to an Of/WN9 type at minimum (e.g., Caputo \& Viotti
1970, Stahl 1986,Viotti 1993).
A similar behaviour was also shown by the Large Magellanic Cloud
star R~127 (e.g. Stahl 1983). These observations suggest that the
variations might take place at constant bolometric luminosity,
implying that the effective stellar radius changes by up to one
order of magnitude between visual minimum and maximum.

Since 2003 GR~290 has been frequently observed with the 30 cm Greve
telescope. On November 13, 2006 it was noticed that the star was
much fainter than previously recorded, with $V$$\simeq$18.4 and
$R$$\simeq$18.5. Thanks to the availability of a 'target of
opportunity service` of the INAF Bologna Astronomical Observatory
(OAB), it was possible to obtain on November 21, 2006 new
photometric and spectroscopic observations of GR~290 with the 1.52~m
Cassini telescope of the Loiano Observing Station of OAB.
These observations confirmed that GR~290 was in a deep minimum
phase. Using the photometric standards given in Table~2 of Viotti et
al. (2006)\footnote{The star named $ku$-$e$ in the table is star "F"
of Kurtev et al. 2001 with a slightly different $B$ value of
17.29$\pm$0.10 than that reported in Table~2 of Viotti et al. The
standard star "E" of Kurtev (2001) has the following data:

ku--e:   $\alpha$=01:35:04.50 $\delta$=+30:42:53.3,  $B$=17.11$\pm$0.10,
$V$=16.44$\pm$0.07, $R$=16.05$\pm$0.05, $I$=15.63$\pm$0.05.
}
we found the following photometric values:
$B$ = 18.46 $\pm$ 0.10 and $V$ = 18.50 $\pm$ 0.04.

%
%==============  FIG. 1 - The spectrum of GR 290  ============
%
 \begin{figure}
 \centering
 \includegraphics[width=16cm]{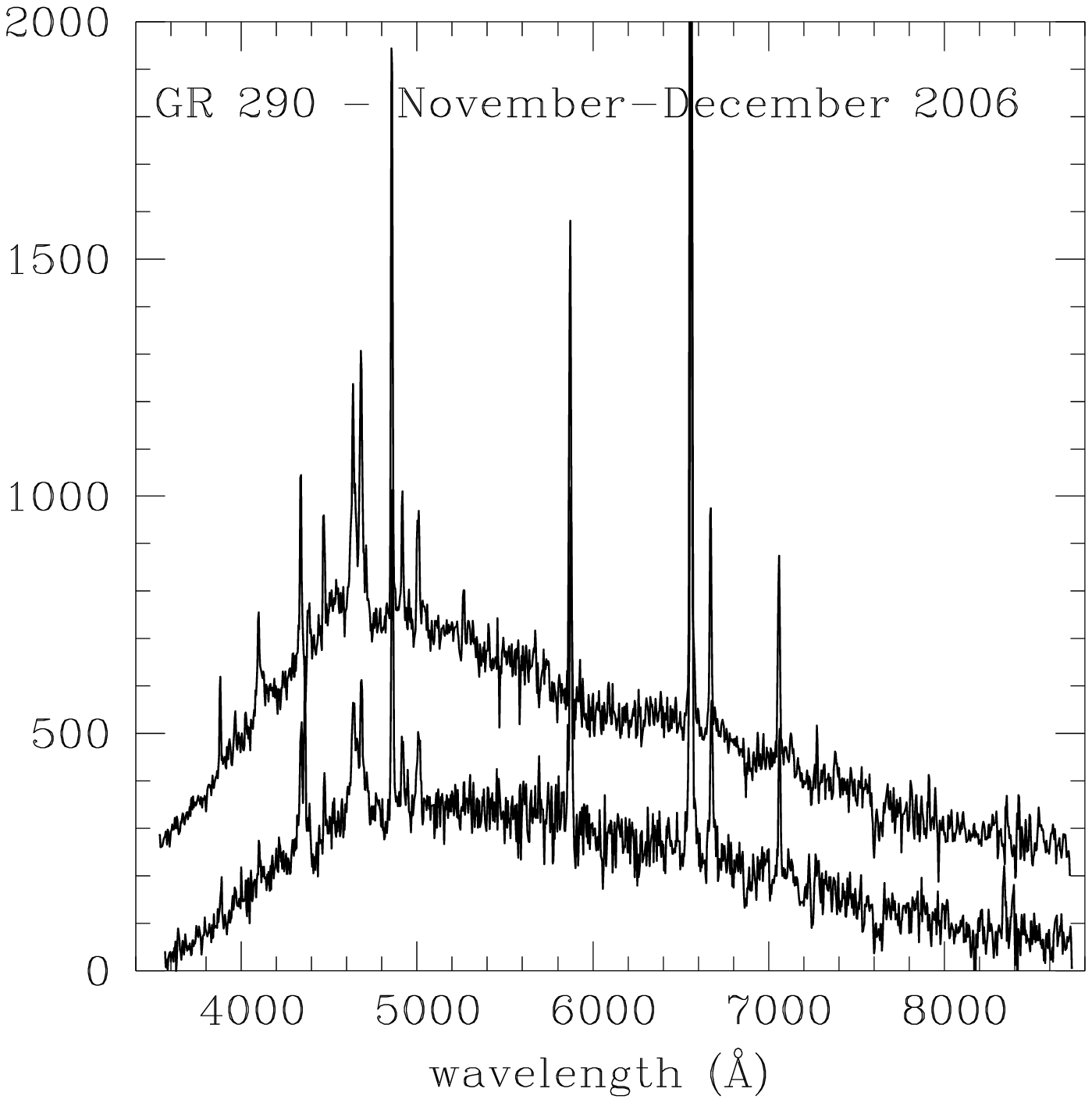}
 \caption{The November (bottom) and December (top) 2006
low resolution spectra of GR~290.
Ordinates are counts with a vertical offset of $+$200
for the December spectrum.
The original November spectrum is slightly smoothed.
The long-wavelength side of the H$\gamma$ emission is affected
by the city light line at 4358 \AA, that is oversubtracted
in the December spectrum.
}
 \label{fig:gr290sp}
 \end{figure}

On the same night we took a low resolution spectrum covering the
range $\sim$3600 \AA\--8800 \AA\ with a dispersion of 3.97 \AA\ per
pixel, and an effective resolution of about 17 \AA. Another spectrum
was obtained with the same setup on December 14, 2006. Because of
the better seeing conditions a smaller entrance aperture was used so
that the effective resolution was of about 13 \AA, with nearly the
same S/N ratio as the November spectrum. Both spectra are shown in
Fig.~1.
The fairly good S/N allows to identify many emission lines, mostly of the Balmer
series and of neutral helium.
Of particular interest is the strong 4640--4713 \AA\ emission feature that is
a blend of N \II\ and N \III\ lines, and of He \II\ 4686 \AA\ and He \I\ 4713 \AA.
In this spectrum the ionised helium line is quite prominent, with an emission peak
reaching 60 $\%$ of the H$\beta$ value.
H$\alpha$ presents asymmetric wings -- stronger on the red side of the line --
that can be attributed to the [N \II] 6548 and 6584 \AA\ emission.

Details on the Loiano and Greve telescopes
are given in \citet{Viotti06}.

%===============   FIG. 2 -- BV LIGHT CURVE OF GR 290 ============
%
 \begin{figure}
 \centering
 \includegraphics[width=16cm]{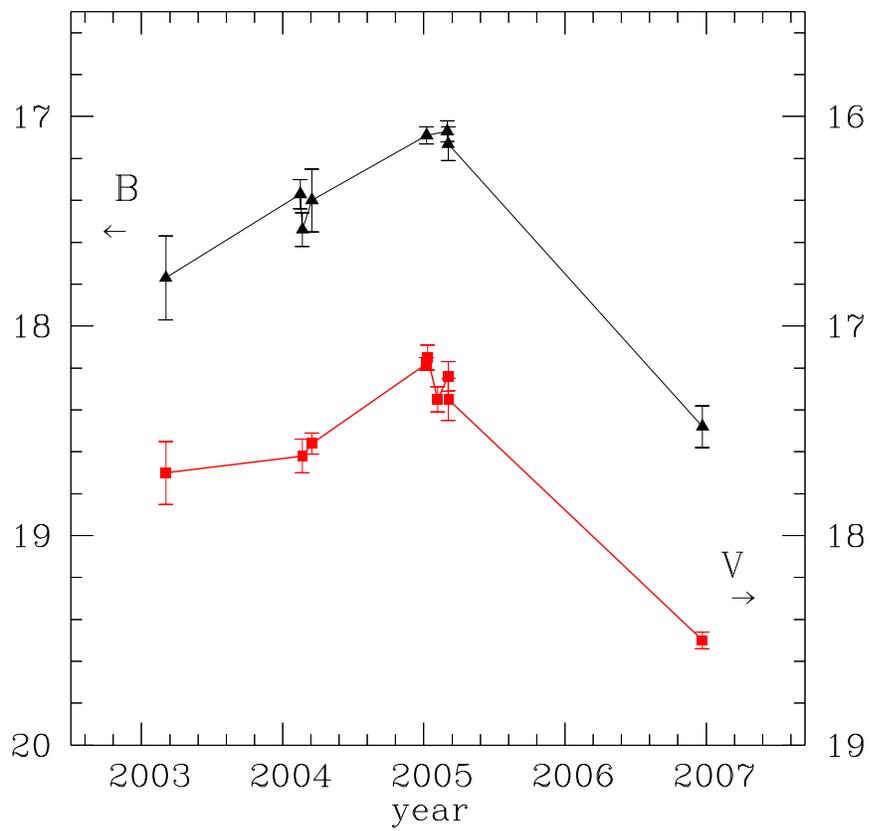}
 \caption{The recent light curve of GR~290 in the $B$ and $V$ bands
(left and right ordinate scales, respectively).
}
  \label{fig:gr290bv}
 \end{figure}
%
%===================================================================

\section{Discussion} \label{sec:discussion}

\subsection{Photometry}

In Fig.~2 we show the recent light curve of GR~290 in the $B$ and
$V$ photometric bands. As noticed by \citet{Viotti06}, during
2003--2004 the star gradually brightened by about a factor of 2
($-$0.7 mag) in all $B$, $V$ and $R$ bands. This brightening phase
was followed by a significant bending of the light curve after
October 2004, suggesting the beginning of a new descending branch.
Our new photometric observations show that the star has become much
less bright with a magnitude difference since the end of 2004 of
+1.4 both in $B$ and in $V$.
{\it This is the deepest minimum of GR~290 in $B$ so far recorded} (see
\citealt{Kurtev01}).
Previously, two minima in the blue were recorded in 1961--62 and
1975--77  with $m_{\rm pg}$=17.6--17.7 (\citealt{Romano78}).
Because of the poor time coverage of the light curve,
other unrecorded minima might have occurred during the past half a century.
The pseudo--period of a few years and the large amplitude of the
recent variation are typical of the
S Dor variable stars (e.g. \citealt{Gend01}).

%
%==============  FIG. 3 - spectral variation  ==============
%
 \begin{figure}
% \centering
 \includegraphics[width=16cm]{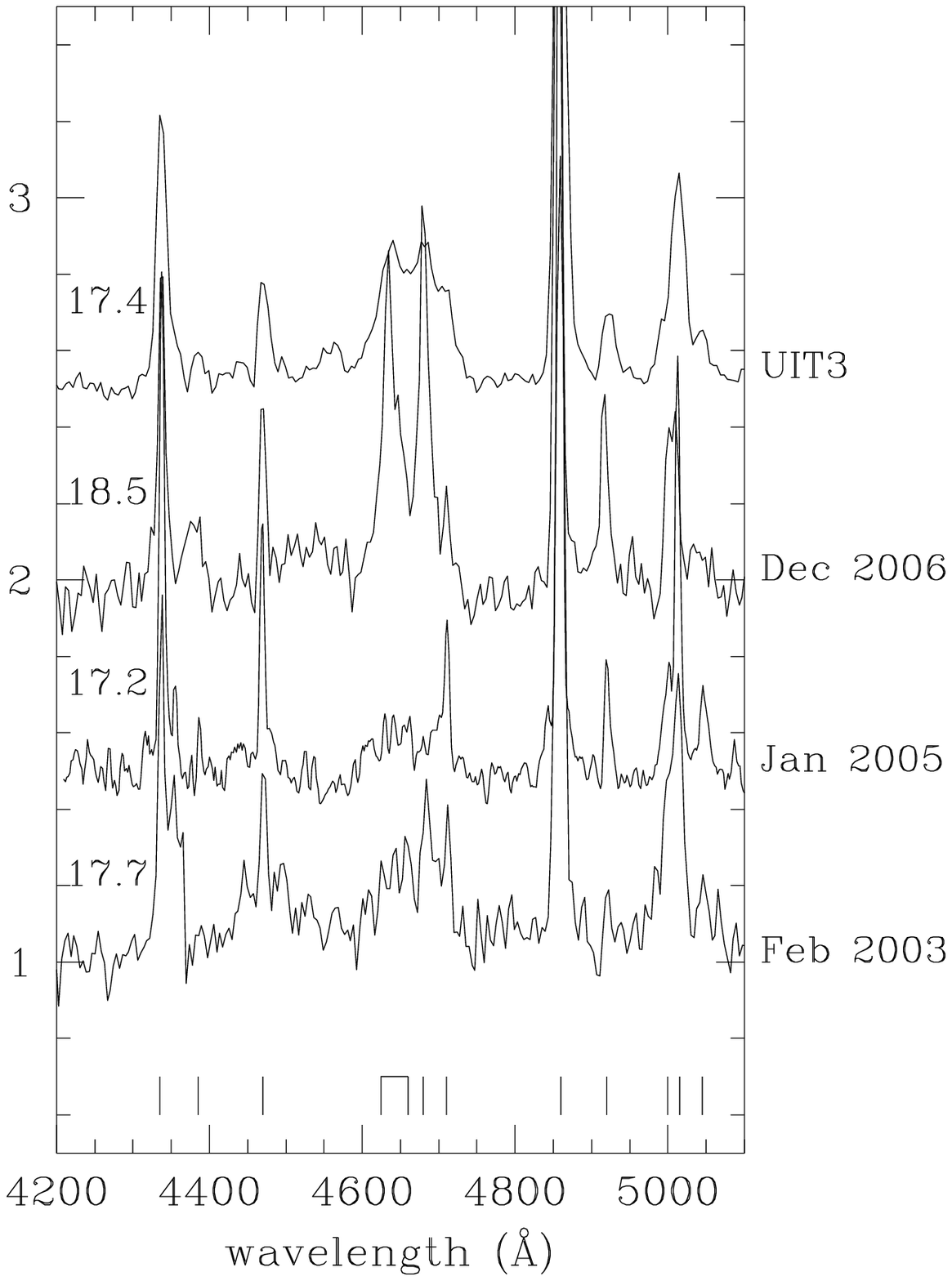}
      \caption{Spectral variation of GR~290 during 2003--2006.
From the bottom: the spectrum of GR 290 on
February 2, 2003, January 13, 2005, December 14, 2006, and the spectrum of
the Of/WN9 star UIT~3 in M~33 of December 7, 2004.
Ordinates are fluxes normalised to the continuum, with vertical offsets.
For each spectrum the visual luminosity is also indicated.
The vertical bars at the bottom mark the following emission
lines: H$\gamma$, HeI 4388 \AA, HeI 4471 \AA, the 4630-4650 \AA\ feature,
HeII 4686 \AA, HeI 4713 \AA, H$\beta$, HeI 4922 \AA, NII
4994--5005 \AA, HeI 5016 \AA, and HeI 5048 \AA.
The peak at 4358 \AA\ is city light.
}
  \label{fig:grspvar}
 \end{figure}
%===========================================================

\subsection{Spectroscopy}

During 2003--2005 many low resolution spectra of GR~290 were
collected by us with the Asiago--Cima Ekar and Loiano Telescopes. As
discussed by \citet{Polcaro03} and by \citet{Viotti06} the spectrum
was always characterised by prominent hydrogen and neutral helium
emission lines, and weak emission of the 4630--4686 \AA\ feature
typical of the Of--type stars. \citet{Viotti06} found that the
latter emissions decreased near the 2004 light maximum. The
4200-5100 \AA\ range of the Loiano December 2006 spectrum,
normalised to the continuum, is compared in Fig.~3  with the spectra
taken previously at Loiano with the same instrumental setup. For
comparison, the figure also includes the Cima Ekar spectrum of the
Ofpe/WN9 star UIT~3 described by \citet{Viotti06}. This spectrum has
a slightly lower spectral resolution, so that the 4630--4660 \AA\
blend, the He \II\ 4686 \AA\ and the He \I\ 4713 \AA\ lines all
appear blended. The other spectra of GR~290 taken at Cima Ekar are
shown in Fig.~7 of \citet{Viotti06}.

In general the strength of the hydrogen and neutral helium emission
lines varied little during 2003--2006 (see also Table~1).
The most
conspicuous variation is shown by the 4630--4686 \AA\ emission blend
which presently is much brighter than in all the previous spectra.
The emission is even stronger than in the M~33 Ofpe/WN9 star UIT~3.
The relative strength of the He \I\ and He \II\ lines, and of
the 4630-4650 \AA\ feature can in principle be used for a tentative spectral
classification of GR~290 during the different epochs, although the
 spectral classification of peculiar, emission line OB stars
is still developing, because of the large variety
of spectra and the very limited number of members,
down to only one representative, for each proposed category.

Using as a reference the Atlas of Peculiar Spectra of OB Stars
(\citealt{Walborn00}), we classify our spectra of GR~290 as (Of)WN11
in 2004--2005 when the 4630--4650 \AA\ blend
and the He \II\ 4686 \AA\ emission lines were very weak, and (Of)WN10
in February 2003 when the He \II\ line is a little stronger than the He \I\ 4713 \AA\ line.
The last November--December 2006 spectra suggest a spectral type (Of)WN9.
The main difference with the spectra of the Atlas is the much
stronger Balmer lines in GR~290 (and in UIT~3 as well).
This might be due to the greater extension of the emitting atmospheric
envelope in these M~33 stars and to the lower optical depth of the hydrogen lines.

Table~1 summarizes the main data on the different phases of GR~290
and reports the equivalent spectral types as discussed above.
The $B$~$-$~$V$ colour index
does not follow the general photometric trend of the S Dor variable stars,
that are bluer when the star is fainter in $V$.
However one should take into account that (as discussed
by \citealt{Viotti06}) in December 2004 GR~290 had
a very blue intrinsic (i.e., reddening--corrected) $B-V$ colour
of $-$0.31.
The flux increase of the 4630--4686 \AA\ emission feature in 2006
might well be accompained by a large deviation of the visual SED
from that of a black--body, and by an increase of the colour index,
such as that observed in GR~290 in 2003 and 2006.
This will require future investigation.

%
%=========  TABLE 1  PHYSICAL PARAMETERS  ===========
%
\begin{table*}
\caption[] {The variable spectrum of GR 290$^1$}
\label{tab:param}
\begin{flushleft}
\begin{tabular}{lcccrrrrr}
\hline
\noalign {\smallskip}
date/target & $V$ & $B-V$ &sp. type$^2$&H$\alpha^3$&H$\beta$&4630-86&4713&5876  \cr
\noalign {\smallskip}
\hline
\noalign {\medskip}
Feb. 02, 2003   & 17.70 & $+$0.07 &WN10eq & 105& 26& 17.3    & 5.4& 25 \\
Feb. 14, 2004   & 17.56 & $-$0.16 &WN11eq & 100& 35&  2.1    & 2.6& 24 \\
Dec. 06-07, 2004& 17.18 & $-$0.09 &WN11eq & 138& 29&  3.9    & 3.1& 27 \\
Jan. 13, 2005   & 17.36 & $-$0.11 &WN11eq & 132& 31&  9.4    & 4.9& 19 \\
Nov.-Dec. 2006  & 18.50 & $-$0.02 &WN9eq  & 143& 32& 40.7$^4$& 2.8& 32 \\
UIT~3           &       & $+$0.01 &Of/WN9 & 127& 34& 31.9    & bl.& 31 \\
\noalign {\smallskip}
\hline
\end{tabular}
\end{flushleft}
{{\it Notes to the table.} ~~
$^1$ Equivalent widths of the emission lines in \AA.
$^2$ Equivalent spectral types for GR 290.
$^3$ Including the line wings and/or the [N \II] lines.
$^4$ The HeII 4686 \AA\ line alone has an
emission equivalent width of 18.4 \AA.
}
\end{table*}
%

%===============   FIG. 4 -- 4650 versus V ============
%
 \begin{figure}
 \centering
 \includegraphics[width=16cm]{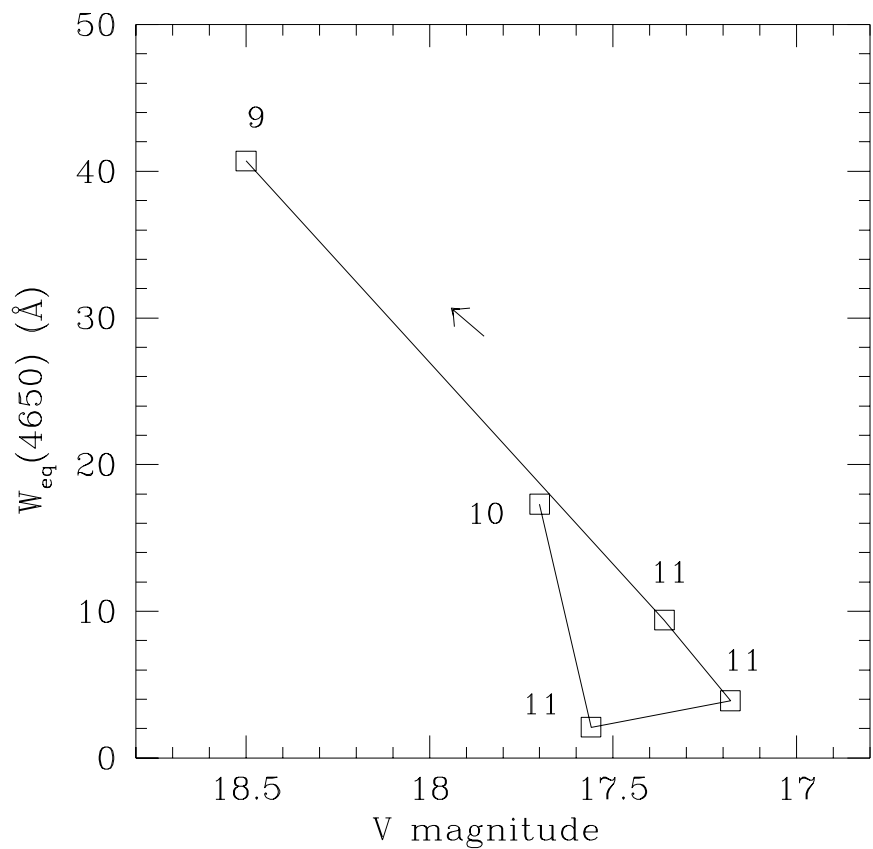}
 \caption{Equivalent width of the 4630-4686 \AA\ emission blend as a
function of the visual magnitude of GR~290 from February 2003
to December 2006. The arrow indicates the direction of the evolution
with time. The labels give the proposed WN-subtypes.
}
  \label{fig:grvar}
 \end{figure}
%
%===================================================================

\section{Conclusions} \label{sec:conclusion}

Viotti et al. (2006) estimated for GR~290 near maximum a bolometric
luminosity of 3$\times$10$^6$ $L_{\odot}$, which places the star
among the brightest LBVs. The 2003--2006 photometric and
spectroscopic monitoring confirmed that the variations are typical
of the S~Doradus variable stars, with the star being hotter when
fainter in the visible. This fact is better illustrated in Fig.~4
where we show the variation of the W$_{eq}$ of the 4630-4686 \AA\
blend as a function of the visual magnitude. The strength of the
blend is linked to the stellar flux in the far ultraviolet, while
the visual magnitude is related to the effective radius and
temperature of the star. The apparent counter--correlation between
visual luminosity and temperature suggests that, as in the best
studied S Dor variable stars, the variations in GR~290 take place at
constant bolometric luminosity.
If this is the case, the star should have increased its effective radius
during 2003--2004, and have reached a maximum radius
(and minimum effective temperature) by the end of 2004.
Since the beginning of 2005 GR~290 has inverted the trend and its
effective radius decreased (by about a factor two),
while the effective (or Zanstra) temperature has become higher.

Of great interest is the fact that GR~290 presently displays
the hottest spectrum ever seen in an LBV.
It would be important to understand whether this result is fortuituous,
because of the poor spectroscopic monitoring of the other LBVs,
or whether it is related to the intrinsic nature of GR~290, which
probably is one of the most massive LBVs.

The similarity of the present spectrum of GR~290
with that of the Of/WN9 stars confirms the link between LBVs and this
category of star, a fact that has to be taken into consideration
in evolutionary models of very massive stars.

Valuable information can be
collected on this most interesting category of stars
by continuous monitoring, even with medium and small size telescopes.

%=====================   ACKNOWLEDGEMENTES =======================

%=====================   BIBLIOGRAPHY   =======================

\end{document}